\documentclass[usenatbib]{mn2e}
\usepackage{epsfig}
\newcommand{\aap}{A\&A}
\newcommand{\aaps}{A\&AS}
\newcommand{\aj}{AJ}
\newcommand{\apj}{ApJ}
\newcommand{\apjl}{ApJL}
\newcommand{\apjs}{ApJS}
\newcommand{\apss}{Ap\&SS}
\newcommand{\araa}{ARA\&A}

\newcommand{\mnras}{MNRAS}

\newcommand{\aapr}{A\&ARv}

\title[Spectroscopy of satellite galaxies]{Spectroscopic confirmation of H$\alpha$-selected 
satellite galaxies}
\author[Clare F. Ivory \& Phil A. James]{Clare F. Ivory$^{1}$\thanks{E-mail:
cfi@astro.livjm.ac.uk} and Phil A. James$^{1}$\\
$^{1}$Astrophysics Research Institute, Liverpool John Moores
  University, Twelve Quays House, Egerton Wharf, Birkenhead, CH41 1LD, UK \\
}
\begin{document}

\date{Accepted 2010 ???. Received 2010 ??? ; in original form 2010 ??? }

%\pagerange{\pageref{firstpage}--\pageref{lastpage}} \pubyear{2010}

\maketitle

\label{firstpage}

\begin{abstract}
%We have spectroscopically tested a method of searching for satellite galaxies around local host galaxies

We present a spectroscopic test confirming the potential of narrow-band optical imaging as a method for detecting star-forming satellites around nearby galaxies. To date the efficiency of such methods, and particularly the fraction of false detections resulting from its use, has not been tested.  In this paper we use optical spectroscopy to verify the nature of objects that are apparently emission-line satellites, taken from imaging presented elsewhere. Observations of 12 probable satellites around 11 host galaxies are presented and used to compare the recession velocities of the host and satellite. This test confirms, in all cases, that there is genuine line emission, that the detected line is H$\alpha$, and that the satellites have similar recession velocities to their hosts with a maximum difference of $\sim250$ km\,s$^{-1}$, consistent with their being gravitationally bound companions. We conclude that the spectroscopy has confirmed that narrow-band imaging through H$\alpha$ filters is a reliable method for detecting genuine, star-forming satellites with low contamination from galaxies seen in projection along the line-of-sight.
\end{abstract}

\section{Introduction}

Satellite galaxies are likely to play an important r\^{o}le in the evolution of galaxies. In recent years the search for representative samples of satellite galaxies has gathered momentum, in a quest to understand their r\^{o}le in the evolution of the present-day galaxy population. In subsequent papers, following the validation tests presented here we will present a study of the luminosities, spatial distributions and star formation (SF) properties of the satellites of nearby bright spiral galaxies, to address two main questions: how many satellites do bright field galaxies have at the current epoch, and can such satellites supply significant amounts of gas to their central galaxies?

These questions are motivated by the influential hierarchical models of galaxy formation and evolution \citep{whit78,fren88}, which are a central component of the lambda cold dark matter ($\Lambda$CDM) cosmological model. The accumulation of matter through the merger of galaxies in hierarchical galaxy formation continues over the age of the universe \citep{sear78} and results in massive galaxies continually accreting gas, stars and dark matter. 

Within this picture, the merging of small satellites with large central galaxies is an important process, as the gas thus supplied may provide the replenishment required for ongoing SF \citep{roch00}  without disk destruction, and may explain the chemical composition observed in the Milky Way \citep{beck04}. The merging of gas-rich satellites either fully or by being stripped of their gas could contribute to the ongoing star forming activity, with or without the disruption of the disk \citep{may06}. Such minor mergers may also play a r\^{o}le in producing `thick disk' components \citep{vela99}, and could contribute to the growth of bulge mass \citep{walk96}. Particularly problematic are the large numbers of field spiral galaxies with little or no bulge component and strong ongoing SF; if these are being built up and supplied with gas through mergers, the infalling galaxies must be small, to avoid disk disruption, and quite numerous. To address these problems, and the general `missing satellites' problem of $\Lambda$CDM \citep{klyp99,moor99}, it is important to have large samples of true satellites, extending to the faintest possible luminosities.

The identification and classification of satellite galaxies was first undertaken using wide-field photographic plates by \cite{hol69}. In this study, a statistical correction was utilised for background galaxies as redshifts were not available.  \cite{zar93,zar97} used multifibre spectroscopic observations to measure the redshifts and quantify the numbers of satellites of field spirals. Whether satellite galaxy mergers can provide gas replenishment of the host galaxies has been investigated through H\,{\sc i} observations \citep{grce09,irwi09} concluding that the current rate of SF in our Galaxy could not be sustained through infall of gas rich satellites alone. However, modelling by \cite{guo08} suggests that minor mergers contribute more to galaxy growth than major mergers. The morphology and chemical abundances of satellites have been studied through broadband photometry \citep{guti04} and spectroscopy \citep{kirb08}. H$\alpha$ observations have been utilised to determine SF rates of 0.01 - 3.7~M$_{\odot}$~yr$^{-1}$, mean value 0.7~M$_{\odot}$~yr$^{-1}$, in satellite galaxies \citep{guti06} selected from Zaritsky's catalogue \citep{zar97}. The highest SF rates appear to be linked to tidal disturbance of the satellites concerned. Further investigation of numbers, spatial distributions, SF and morphology of bright satellites has been undertaken using the SDSS data \citep{ann08,azza07,bail08,bosc08,chen08,chen06,yang06}, and observational predictions modelled using the Millennium Simulation \citep{sal07}. These studies show indications of morphological transformation of satellites from late- to early-types resulting from their proximity to central galaxies, and the consequent removal of gas reservoirs (`strangulation').

In addition to the r\^{o}le satellite galaxies may play in the gas replenishment of their host galaxies, they are also important to the investigation of mass distributions at large radii in dark matter haloes \citep{prad04,mado04}. 

There have been many dwarf galaxies observed in the extended Galactic halo, the most massive and gas-rich being the Large and Small Magellanic Clouds, from which the Magellanic Stream emanates.  This stream would indicate the early stages of gas exchange \citep{moor94}. For the infall of gas from satellite galaxies to be a realistic resolution to the gas replenishment problem, gas-rich and hence star-forming satellites should be common near bright star-forming disk galaxies (\citealt{sanc08}, and references therein).

\begin{figure}

\includegraphics[width=82mm]{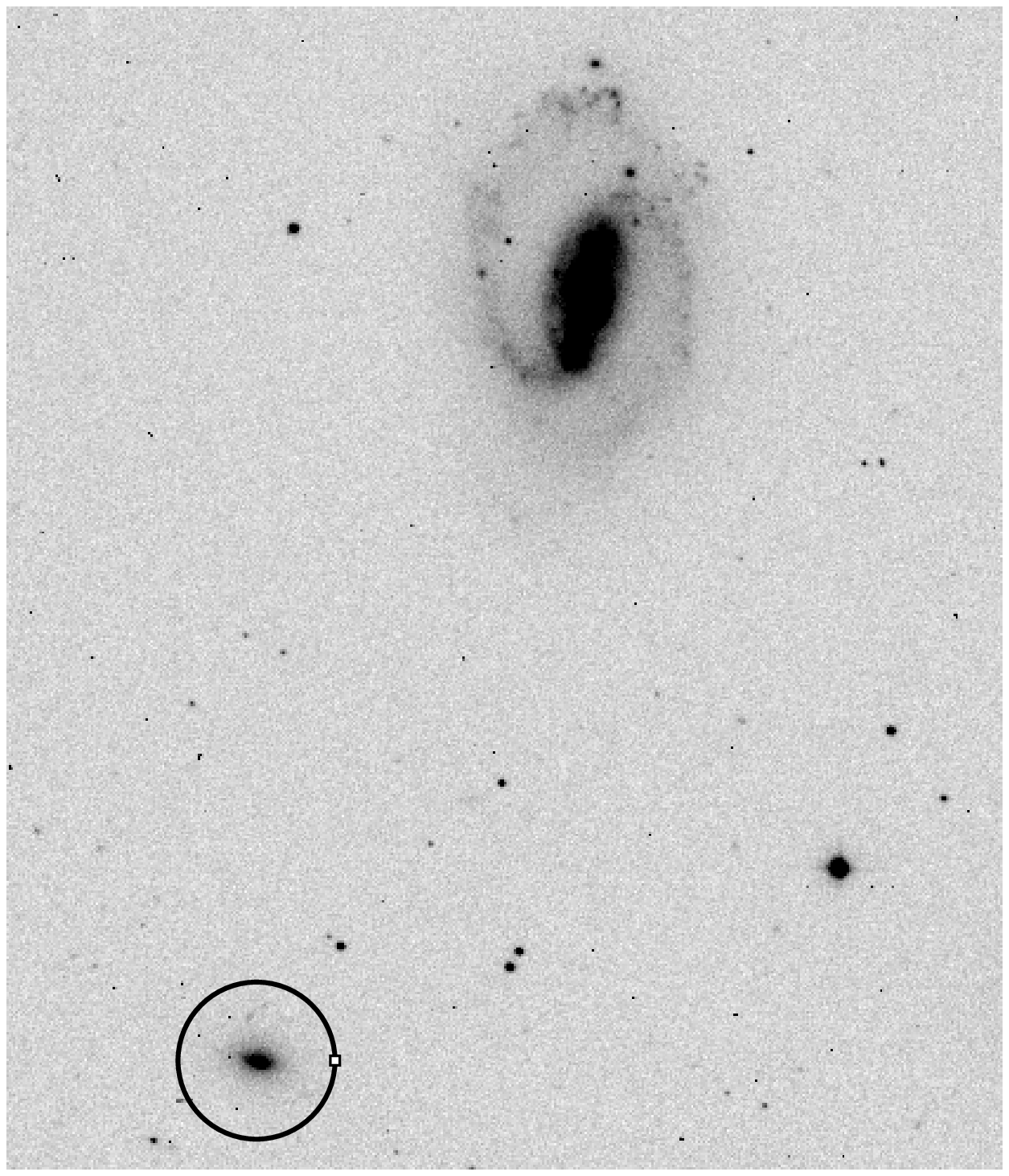}
\includegraphics[width=82mm]{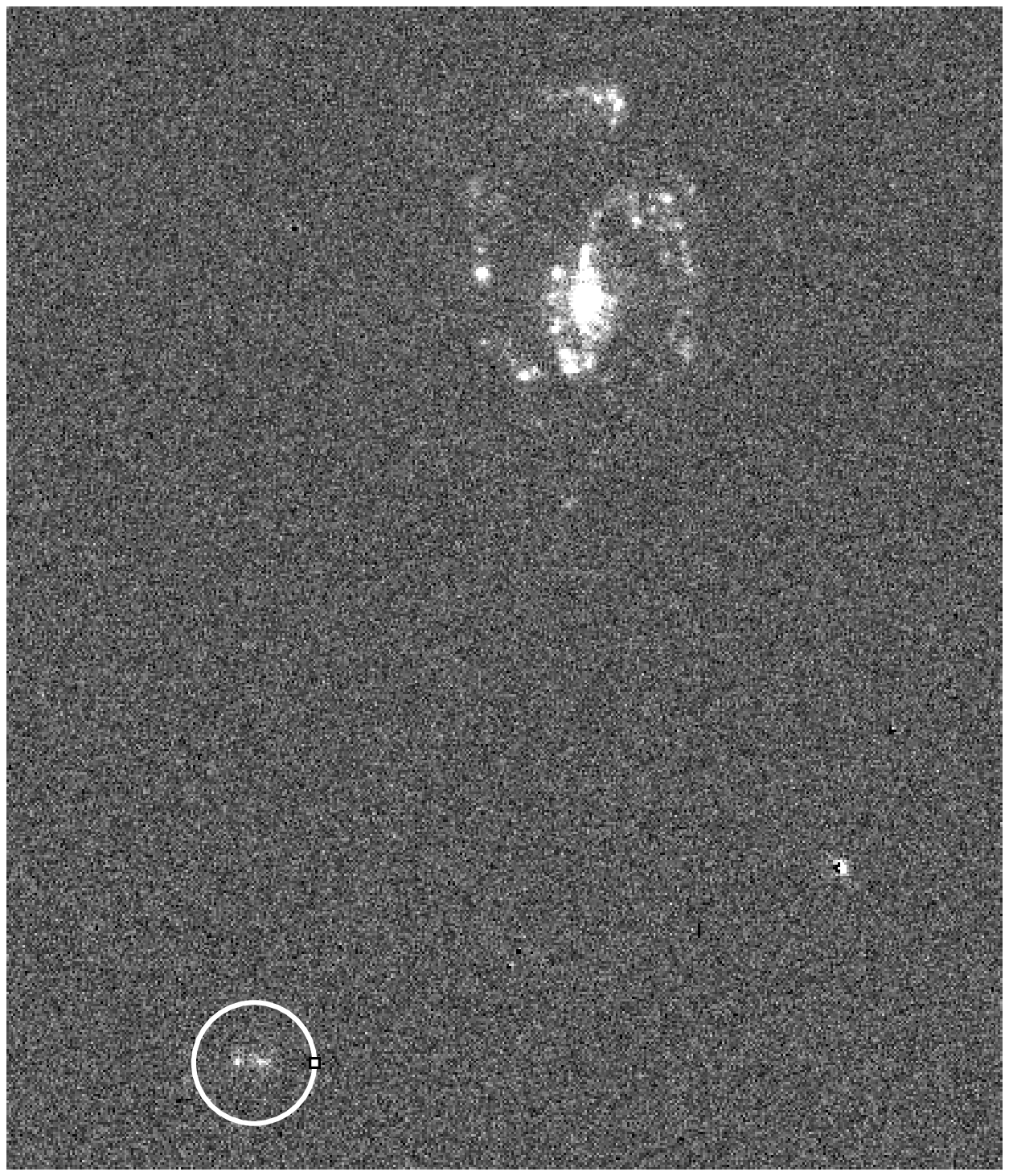}

 \caption{UGC4574 - $R$-band (top) and continuum-subtracted H$\alpha$ (bottom) $\sim$\,4$\times$5 arcmin images. The satellite galaxy is circled}.
 \label{imagefig}
\end{figure}

H$\alpha$ imaging as a method for satellite detection has the advantage that a wide area is covered with one observation, and any objects detected must have significant line emission within the appropriate passband, which excludes most of the projected line of sight companions. Additionally, H\,{\sc ii} emission tends to have a high surface brightness even in low surface brightness regions of massive spiral galaxies and dwarf galaxies, and therefore faint companions can be observed with moderate integration times and moderate-aperture telescopes. Thus, this is potentially a sensitive and quick method for detecting even the very low-mass star-forming satellites of low-redshift galaxies.

The H$\alpha$ narrow-band imaging technique has been extensively employed to study SF in individually targeted bright field galaxies e.g. \cite{ken83}, as reviewed by \cite{ken98}, and galaxies in low-redshift clusters \citep{bos02a,bos02b,dav02,igl02,koo01,koo04b,koo04a}.  In addition, H$\alpha$ emission has been used to search for emission line galaxies in objective prism based searches for galaxies in both the cluster environment \citep{mos88,mos93,ben98} and over field areas \citep{zam90,vit96a,vit96b,alo99,sal00,sal05}. However, none of these studies has either searched for, or analysed the properties of, satellite or companion galaxies. Thus our programme is, to the best of our knowledge, the first of its type.

The measurement of H$\alpha$ narrow band emission in optical imaging requires the removal of  continuum emission, which can be done through the subtraction of appropriately-scaled $R$-band images. There are uncertainties in this process which may lead the observer to question whether there is H$\alpha$ emission, or whether the apparent emission is a result of poor continuum subtraction, particularly from poorly subtracted stars or other image artifacts. Most of these can be rejected from visual inspection of the continuum-subtracted H$\alpha$ image (an example of a stellar residual is shown towards the lower-right hand corner of the example image presented in Fig.~\ref{imagefig}, with characteristic positive and negative zones), and through their low equivalent widths.  However, at faint magnitudes and for stars with very extreme colours this process is less clear-cut. 

There is also the possibility of other, shorter wavelength emission lines from intermediate-redshift galaxies, for example the {[O\,{\sc iii}]}\,$\lambda$5007\,{\AA} line, being redshifted into the passband of the H$\alpha$ filter. Additionally, the narrow-band filter widths, typically 1$\%$, allow H$\alpha$ emission to be detected in a range which is much wider than the velocities of the satellites with reference to their hosts.  This means that even if the line detected is truly H$\alpha$, any apparent satellites could easily be separated by $\sim1000$ km\,s$^{-1}$ from the central galaxy, and thus not be truly associated companions.

The aims of this paper are to analyse spectroscopically the potential satellites detected through H$\alpha$ imaging, to quantify the above uncertainties and ascertain the reliability of the H$\alpha$ imaging method for detecting true companions of field spiral galaxies.

From an original imaging data set of 262 field spiral galaxies, 73 candidate satellites associated with 56 host galaxies were discovered of which 11 hosts and 12 candidate satellites are to be spectroscopically tested and results presented in the present paper. The complete set of imaging data are to be analysed to determine the integrated properties and spatial distribution of SF, the SF rates and timescales, and their $R$-band and H$\alpha$ luminosities. These properties of the central galaxies, and analyis of the numbers and properties of their satellites, will be studied in a series of separate papers.

The original imaging data set of nearby spiral galaxies was observed with the Jacobus Kapteyn Telescope (JKT) for the H$\alpha$ Galaxy Survey, H$\alpha$GS \citep{jam04}. Additional galaxies were observed with the Isaac Newton Telescope (INT) for a study of supernova-hosting galaxies \citep{and08}. The recession velocities obtained from the NASA/IPAC Extragalactic Database (NED) for the central galaxies discussed in the present paper lie in the range 1500 - 6000 km\,s$^{-1}$, and all had candidate satellites that were apparent in the continuum-subtracted H$\alpha$ images. An example galaxy and satellite from this imaging data set is shown in Fig.~\ref{imagefig}; here the satellite can been seen in the bottom left hand corner of both the $R$-band and H$\alpha$ images.

%\begin{figure}

%\includegraphics[width=85mm]{4574rbwblackcirc.ps}
%\includegraphics[width=85mm]{4574hbwwtcirc.ps}

 %\caption{UGC4574 - $R$-band (top) and continuum-subtracted H$\alpha$ (bottom) $\sim$\,4$\times$5 arcmin images. {\bf The satellite galaxy is circled}.}
 %\label{imagefig}
%\end{figure}

\section{Observations}

The optical galaxy spectra presented in this paper were observed between the 4th and 8th January 2009 with the INT on La Palma. The observations were made using the Intermediate Dispersion Spectrograph (IDS), with the R600R grating, the 235-mm camera and the EEV10 CCD. The spectral range was 5900\,{\AA} - 8600\,{\AA} to ensure the H$\alpha$ emission line lies within the central unvignetted region of the CCD. The slit width used was 2.0 arcseconds and was positioned centrally on the visual peak of each central and satellite galaxy observed, using the image on the INT acquisition TV screen, with a slit position angle of 0$^{\circ}$ (N-S orientation) for all spectra. Wavelength calibration was achieved by taking neon lamp spectra throughout the night after each galaxy observation. Eleven host galaxies and twelve candidate satellites were observed. Flux calibrations were not required and therefore standard stars were not observed. Integration times varied between 300 and 1200 seconds. 
The frames were debiased and flat-fielded, cosmic rays and sky lines were removed, and the extracted spectra were wavelength-calibrated, using $Starlink$ software.

\begin{figure*}
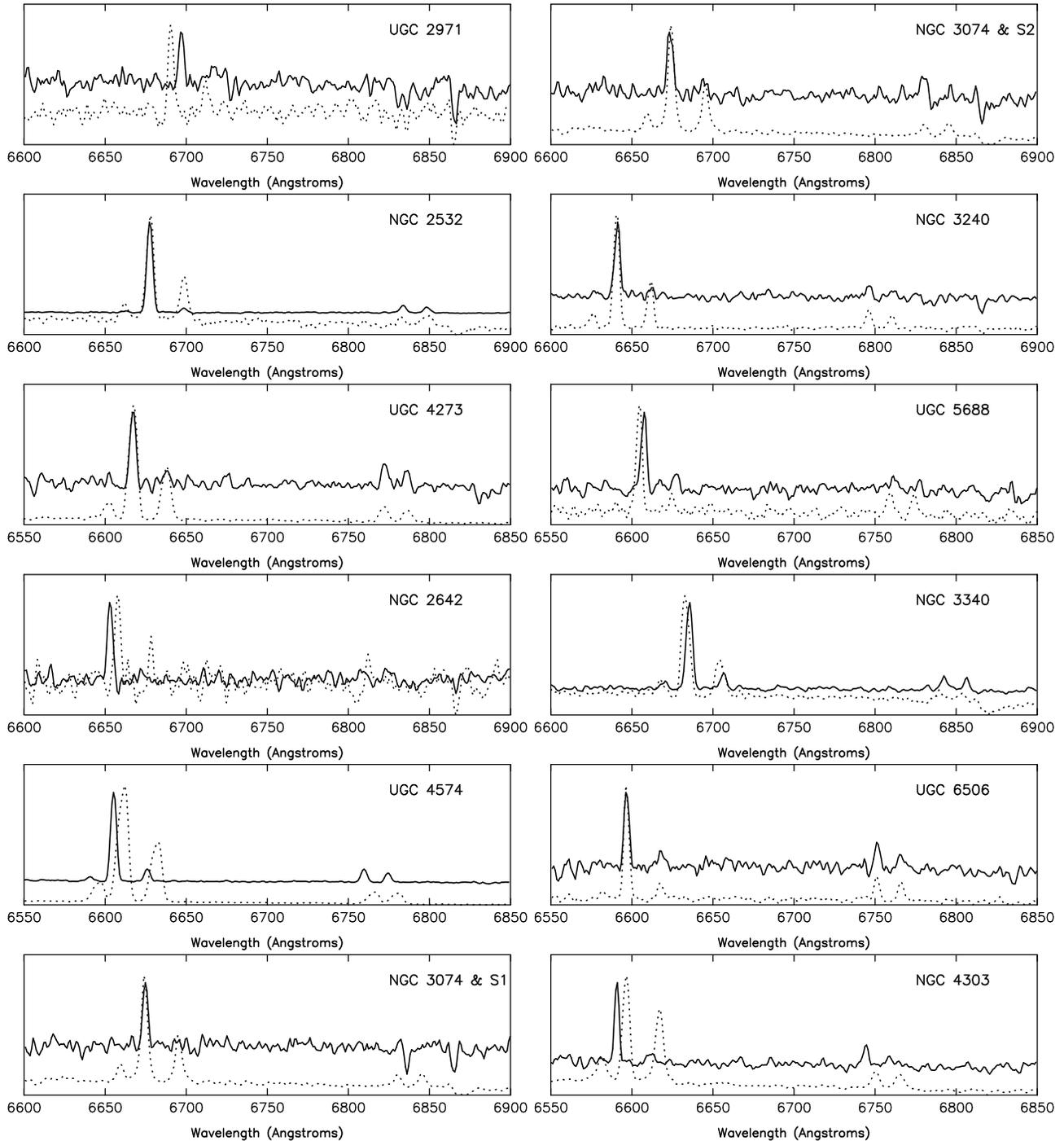

\includegraphics[angle=-90,width=85mm]{u2971_spec.ps}
\includegraphics[angle=-90,width=85mm]{n3074_2_spec.ps}
\includegraphics[angle=-90,width=85mm]{n2532_spec.ps}
\includegraphics[angle=-90,width=85mm]{n3240_spec.ps}
\includegraphics[angle=-90,width=85mm]{u4273_spec.ps}
\includegraphics[angle=-90,width=85mm]{u5688_spec.ps}
\includegraphics[angle=-90,width=85mm]{n2642_spec.ps}
\includegraphics[angle=-90,width=85mm]{n3340_spec.ps}
\includegraphics[angle=-90,width=85mm]{u4574_spec.ps}
\includegraphics[angle=-90,width=85mm]{u6506_spec.ps}
\includegraphics[angle=-90,width=85mm]{n3074_1_spec.ps}
\includegraphics[angle=-90,width=85mm]{n4303_spec.ps}

 \caption{The spectra of the galaxy hosts (dotted line) and their satellites (solid line). The flux scales are arbitrary and therefore not shown. The satellite spectrum is offset vertically for clarity.  H$\alpha$, {[N\,{\sc ii}]}\, and {[S\,{\sc ii}]}\, lines are evident in most spectra.}
\label{specfig}
\end{figure*}

\section{Results}

\subsection{H$\alpha$-derived galaxy recession velocities}

The galaxy host and satellite spectrum pairs are shown in Fig.~\ref{specfig}.  In each case, the satellite spectrum is shown as the solid line, and the satellite spectrum is offset vertically by an arbitrary amount for ease of comparison.  The purpose of these spectra is to confirm the strongest line as H$\alpha$ by identifying other lines in the spectra, where available; and to get recession velocities from accurate determination of wavelengths of emission lines.
The latter were measured by fitting a Gaussian profile to each wavelength-calibrated spectrum. These were converted to recession velocities assuming the rest wavelength of H$\alpha$ to be 6562.8\,{\AA}.

\begin{table*}
\caption{Spectroscopically determined host and satellite recession velocities}
\begin{tabular}{|c|c|c|c|c|c|c|c|c|} 
\hline
Host Galaxy &\multicolumn{2}{|c|}{Satellite Position Equ.J2000} & Separation & Host cz (NED) & Host cz & Satellite cz & $\Delta$ cz & $v_{rot}$ \\
& RA & Dec & kpc & km\,s$^{-1}$ & km\,s$^{-1}$ & km\,s$^{-1}$ & km\,s$^{-1}$ & km\,s$^{-1}$ \\  \hline
 UGC 2971 & 04$^{h}$09$^{m}$26$^{s}\!\!.$78 & +36$^{\rm o}$59$^{\prime}$44$^{\prime\prime}\!\!.$17 & \phantom{0}81.6 & 5881 $\pm$ 5 & 5934 $\pm$ 14  & 6153 $\pm$ 14 & 219 $\pm$ 20 & 195 \\ 
 NGC 2532 & 08 10 25.20  & +34 00 18.38  & \phantom{0}75.4 & 5260 $\pm$ 4 & 5256 $\pm$ \phantom{0}5 &  5240 $\pm$ \phantom{0}4 & \phantom{0}16 $\pm$ \phantom{0}6 & 257\\ 
 UGC 4273 & 08 12 58.50 & +36 11 53.25  & \phantom{0}35.5 & 2471 $\pm$ 10 & 2478 $\pm$ \phantom{0}9 & 2496 $\pm$ \phantom{0}6 & \phantom{0}18 $\pm$  11 & 174 \\
 NGC 2642 & 08 40 34.14  & --04 09 33.24  & \phantom{0}53.1 & 4345 $\pm$ 8 & 4325 $\pm$ 14 & 4112 $\pm$ 15 & 213  $\pm$  21 & 250\\  
 UGC 4574 &  08 48 23.35  & +74 02 15.61  & \phantom{0}38.6 & 2160 $\pm$ 7 & 2200 $\pm$ \phantom{0}8 & 1944 $\pm$ \phantom{0}5 & 256  $\pm$ \phantom{0}9 & 189\\  
 NGC 3074 & 09 59 48.80  & +35 23 38.91  & \phantom{0}33.1 & 5144 $\pm$ 5 & 5087 $\pm$ 19 & 5124 $\pm$ 18 & \phantom{0}37 $\pm$ 26 & 230\\
  $\prime\prime$ & 10 00 00.59  & +35 21 07.88  & \phantom{0}98.5 & $\prime\prime$  &  $\prime\prime$  &  5056 $\pm$ 13 & \phantom{0}31 $\pm$ 23 & $\prime\prime$ \\
 NGC 3240 &  10 24 23.93  & --21 49 10.01  & \phantom{0}29.3 & 3550 $\pm$ 9 & 3529 $\pm$ 11 & 3583 $\pm$ \phantom{0}3 & \phantom{0}54 $\pm$ 11 & 151 \\
 UGC 5688 & 10 30 39.34  & +70 01 32.79  & \phantom{0}18.0 & 1921 $\pm$ 6 & 1896 $\pm$ \phantom{0}4 & 2034 $\pm$ \phantom{0}9 & 138  $\pm$  10 & 102\\
 NGC 3340 & 10 42 42.72  & --00 20 52.17  & 142.6 & 5558 $\pm$ 6 & 5475 $\pm$ \phantom{0}5 & 5614 $\pm$ \phantom{0}4 & 139  $\pm$ \phantom{0}6 & 206\\
 UGC 6506 & 11 31 19.72  & +28 31 21.14  & \phantom{0}26.9 & 1580 $\pm$ 3 & 1546 $\pm$ \phantom{0}5 &  1553 $\pm$ \phantom{0}7 &\phantom{0}\phantom{0}7  $\pm$ \phantom{0}9 & \phantom{0}67 \\
 NGC 4303 & 12 22 27.21  & +04 33 58.60  & \phantom{0}42.8 & 1567 $\pm$ 5 & 1526 $\pm$ 10 & 1263 $\pm$ \phantom{0}8 & 263  $\pm$  13 & 206\\
\hline
\end{tabular}
\label{table2}
\end{table*}

Uncertainties in the neon arc calibration had rms values between 0.01\,{\AA} and 0.02\,{\AA}. Wavelength uncertainties were further quantified by repeated Gaussian fits to each H$\alpha$ emission line, the estimated error on each fit varied between 0.06\,{\AA} and 0.4\,{\AA}  and, where there were multiple spectra of a galaxy, the central wavelength of the peak typically varied by 0.1\,{\AA} to 0.2\,{\AA}. Uncertainties in individual central wavelengths were taken as the mean error determined from the Gaussian fits, as this was at least a factor of 2 larger than any other source of uncertainty and it implicitly incorporates the effects of varying signal-to-noise ratio from pixel to pixel.

\begin{figure}
\includegraphics[angle=00,scale=0.15,width=85mm]{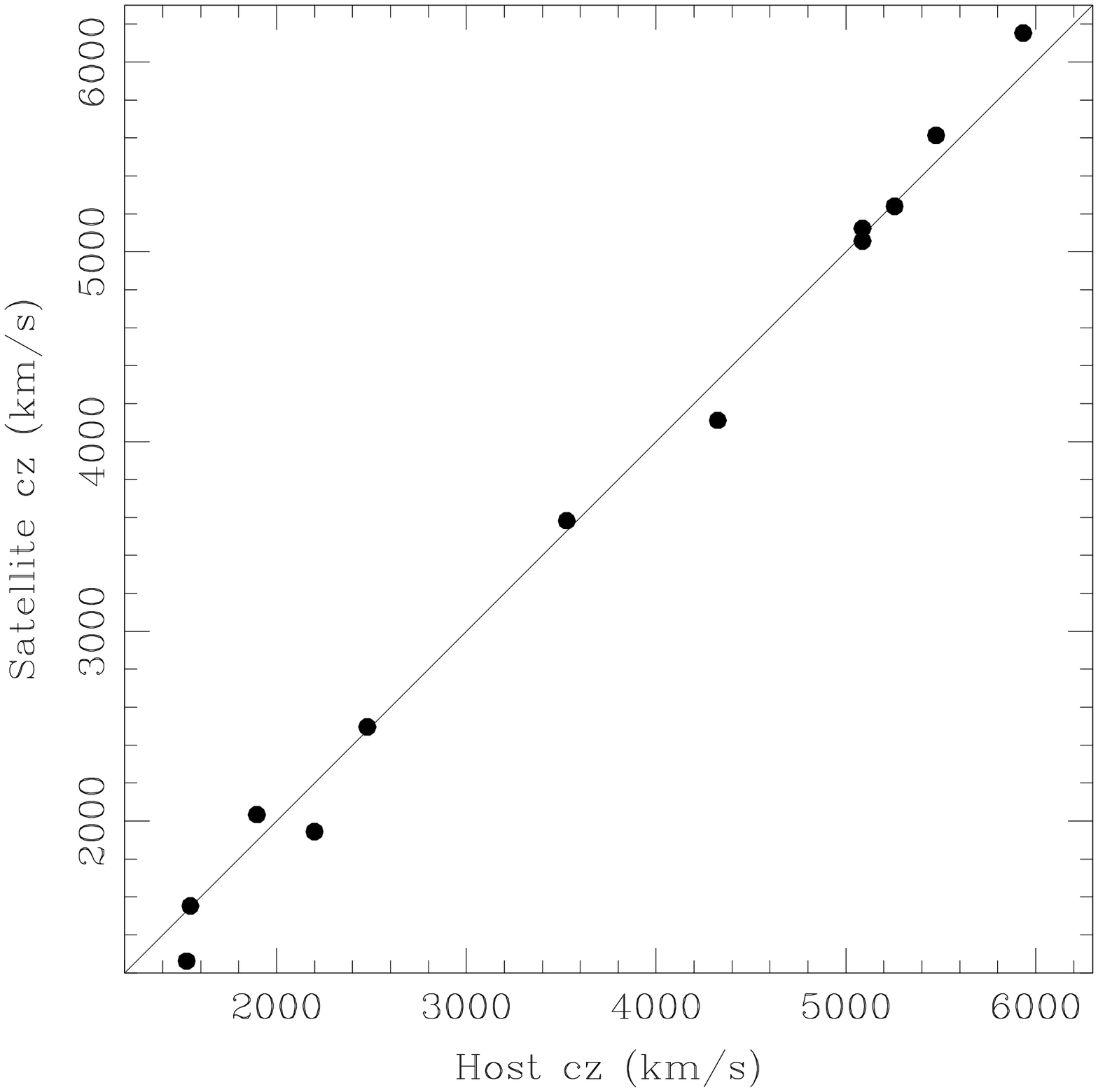}
 \vspace{0cm}
 \caption{Host and satellite recession velocities. The solid line indicates the position of equal host and satellite velocity.}
 \label{velplot}
\end{figure}

\subsection{Testing the nature of the candidate satellites}

Projected separations between hosts and satellite galaxies were calculated by assuming both to lie at distances corrected for Virgo Infall only, taken from NED. The results are shown in Table \ref{table2}, which lists central galaxy catalogued names, positions, satellite and host projected separations, host and satellite heliocentric recession velocities and velocity differences (with the velocities in columns 6 to 8 all being derived from the data presented here), and in the final column estimates of the central galaxies' circular rotation velocities (discussed below). The recession velocities measured from our spectroscopy (col. 6) can be seen to be  in good agreement with the heliocentric recession velocities taken from NED (col. 5), with maximum differences within a few tens of km~s$^{-1}$.  While these differences are larger than the formal errors in some cases, they are in every case smaller than the internal kinematic velocities. In every case, the quoted literature value is derived from H{\sc i} 21~cm observations to minimise extinction-related uncertainties.

The velocities determined for each host galaxy and associated satellite are plotted against each other in Fig.~\ref{velplot}. The plot demonstrates the close correlation in recession velocities between the host and candidate satellite galaxies and that the velocity differences are relatively small.

The initial concern when searching the H$\alpha$ continuum subtracted imaging data was the possibility of mistaking reduction artifacts for satellite galaxies, whereby poorly subtracted stars or other bright foreground or background objects could be erroneously interpreted as line emission. However, all of the candidate  satellites spectroscopically observed had clear emission lines in their spectra, thus demonstrating that continuum subtraction errors in our imaging are not causing problems in identifying true emission.

The second uncertainty was whether shorter wavelength lines other than H$\alpha$, e.g. the {[O\,{\sc iii}]\,$\lambda$5007} forbidden line, had been redshifted into the passband. However, when studying the spectra in every case evidence was found for the line being H$\alpha$, with this being confirmed in most spectra by the presence of at least the stronger of the nearby [N{\sc ii}] lines, and in many cases the longer-wavelength [S\,{\sc ii}] lines.  When analysing the satellite spectra (solid lines in Fig.~\ref{specfig}), the emission line identifications are as follows (starting with the most secure): [N{\sc ii}] and [S\,{\sc ii}] lines are seen from the satellites of UGC 4574, NGC 3340, NGC 2532, and probably from the satellites of UGC 4273, NGC 3240, UGC 5688, UGC 6506, NGC 2642 and NGC 4303; the satellite of UGC 2971 and NGC 3074 satellite number 2 possibly show the [N{\sc ii}]\,$\lambda$6584\,{\AA} line.  In addition to the confirmation from separate line identifications, it should be noted that both satellites of NGC 3074, and the satellite of UGC 6506 have their strongest lines very closely coincident with H$\alpha$ from the central galaxy, and subsequent to these observations we found a literature redshift for the satellites of UGC 2971and NGC 4303, in good agreement with the redshift obtained by assuming the strong lines to be H$\alpha$.   

\begin{figure}
\includegraphics[angle=00,scale=0.15,width=85mm]{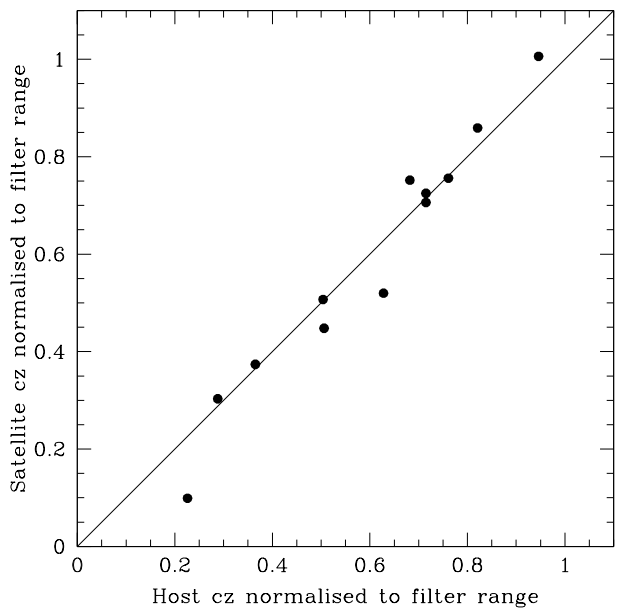}
 \vspace{0cm}
 \caption{Host and satellite recession velocities, normalised to the velocity range allowed by the relevant filter passbands. The solid line indicates the line of equality. The formal error bars are smaller than the points.}
\label{filtplot}
\end{figure}

The narrow-band H$\alpha$ filters used for imaging have widths of $\Delta\lambda$/$\lambda_0$ $\simeq$ 1$\%$ allowing for a range of galaxy recession velocities, within the full-width half maximum of the transmission curve, that can be as large as 6130 km\,s$^{-1}$.  Therefore the method of identifying satellites by H$\alpha$ imaging may include apparent associations, which are line-of-sight star forming dwarf galaxies within this velocity range, but which could have a significant recession velocity difference and thus not be truly associated with the putative host galaxy. The strong correlation between host and satellite recession velocities shown in Fig.~\ref{velplot}. provides some evidence that this is not a major problem, but there is a significant selection effect which undermines the use of Fig.~\ref{velplot} in this context. This effect results from the use of narrow-band imaging to select the satellites plotted in this figure, which therefore requires them to have recession velocities fairly close to those of the central galaxies. Thus, even if the apparent companions were an unassociated field population, this passband selection effect would force a general trend in Fig.~\ref{velplot}.

To overcome this effect, we have constructed Fig.~\ref{filtplot} where the fitted wavelengths have been converted to a relative wavelength, normalised to the range permitted by the narrow-band filters used in the imaging observations. In this plot, zero corresponds to the H$\alpha$ line lying at the blue limit of the appropriate filter (defined as where the filter transmission has fallen to 50\% of the peak value) and 1 corresponds to the equivalent point at the red limit of the filter. With this parametrisation, the apparent satellites could lie anywhere in the $y$ direction between 0 and 1 if they were just random field objects, but in fact this plot shows that the wavelength offset of the candidate satellites is significantly less than that allowed by the wavelength selection of the filters. The Pearson correlation coefficient is 0.979, and for 12 points a correlation coefficient this high has a $10^{-11}$ probability of occurring by chance.  This demonstrates that the candidate satellites are highly unlikely to be a random line-of-sight associations.

A simple calculation was undertaken to determine whether we might have expected to detect any line-of-sight companions in Fig. \ref{filtplot}.  This was based on an estimate of the space density of faint star-forming galaxies from the H$\alpha$ luminosity function (LF) of \cite{jamknap08}.  Taking the H$\alpha$ LF from the luminosity of the LMC down to 3\% of the LMC, we estimated there to be $\sim$\,0.076 such galaxies per cubic Mpc.  The lower limit in this calculation is based on the estimated sensitivity limit of this method, as discussed in \citet{jam08}. We then calculated the volume surveyed by each of the imaging exposures from which the satellite catalogue was derived, taking account of the area covered by the CCD and the radial range over which the H$\alpha$ line lies within the bandpass of the filters used.  For the JKT imaging, we find that we would expect typically one interloper per 55 image fields, whereas for the INT imaging (with a wider field and broader narrow-band filters) we expect approximately one interloper per 10 imaged fields.  These are lower by a factor of a few than the rates at which we have found apparent companions (approximately 1 per 10 JKT fields, \cite{jam08}; and 1 per 3 INT fields, James \& Ivory in preparation). Thus it is to be expected that most companions are truly associated with the imaged central galaxies, but it is somewhat surprising that none of the 12 in the present study appear to be an interloper.

We next determine whether the differences in recession velocity are consistent with the satellites being gravitationally bound to their host galaxies. If we assume the total masses of the host galaxies to be in the range 10$^{11}$ - 10$^{12}$\,{M$_\odot$}, the predicted circular orbital speed of a satellite with the mean observed separation of 56 kpc would be 88 - 277 km\,s$^{-1}$. This is consistent with the mean of the difference in the recession velocities which is 116 km\,s$^{-1}$, even allowing for the average effects of inclination and orbital phase. 

Alternatively, we can compare the velocity differences with the rotational velocities expected for material in the outer disks of galaxies, which may be a predictor of satellite velocities under the (major) assumption that rotation curves are flat and extend to the radii probed by these satellites.  At the simplest level, bright field spiral galaxies are generally found to have rotational velocities of $\sim$ 200 km\,s$^{-1} $ \citep{vogt04, frid05}, consistent with the range of central-satellite velocity differences we observe. At a slightly higher level of sophistication, rotational velocities can be predicted for individual galaxies from our measured total $R$-band magnitudes (taken from the imaging data, Ivory et al. 2010 in preparation), using the Tully-Fisher (TF) relation for field galaxies. The $R$-band TF relation of \citet{pier92} gives values of $v_{rot}$ between 70 and 260~km~s$^{-1}$, as listed in the final column of Table 2. In most cases (8/12) the host-satellite velocity differences are less than the TF predicted rotation velocities, i.e. completely consistent with the satellites being on bound orbits. In the other 4 cases, the velocity differences are only 12--35\% greater than predicted rotational velocity, consistent with the uncertainties of the TF relation, and if a direct radial infall is assumed rather than a circular orbit \citep{hop08}, the expected relative velocities would be larger than for circular orbits. Projection effects could be responsible for the galaxies with smaller velocity differences. Overall, we conclude that the observed velocity differences are consistent with all of the candidate satellite galaxies being bound to the central galaxies.

The close correlation of the host and satellite galaxy recession velocities confirms that H$\alpha$ imaging finds true companions. The satellites and hosts were always found to have the H$\alpha$ emission line and therefore the continuum subtraction for the imaging data successfully isolates this recombination line. The spectra also clearly identify the H$\alpha$+{[N\,{\sc ii}]} lines, and often the S\,{\sc ii} lines, and therefore there have been no cases of {[O\,{\sc iii}]\ $\lambda$5007} or other shorter wavelength lines from line-of-sight background galaxies.

\begin{table}
\caption{Satellite galaxy luminosities and SF rates}
\begin{tabular}{|l|c|c|c|c|} 
\hline

Host Galaxy & $m_{R}$ & $L_{R}$ & SFR & EW(H$\alpha$) \\
\hline
            & Mag    & L$_{\odot}$ & M$_{\odot}$~yr$^{-1}$ & \AA \\
\hline
UGC 2971 &	14.05 &	9.12$\times$10$^9$ &	0.510 &		7.6\\
NGC 2532 &	16.02 &	1.20$\times$10$^9$ &	0.767 &		87.3\\
UGC 4273 &	15.26 &	5.81$\times$10$^8$ &       0.018 &	        8.2\\
NGC 2642 &	15.00 &	2.08$\times$10$^9$ &	0.514 &		33.5\\
UGC 4574 &	14.55 &	8.55$\times$10$^8$ &       0.058 &	        19.0\\
NGC 3074(S1) &	17.71 &	2.39$\times$10$^8$ &	0.021 &		11.6\\
NGC 3074(S2) &	16.10 &	1.05$\times$10$^9$ &	0.082 &		10.5\\
NGC 3240 &	16.45 &	3.66$\times$10$^8$ &	0.035 &		12.8\\
UGC 5688 &	15.03 &	4.81$\times$10$^8$ &	0.097 &		69.8\\
NGC 3340 &	15.17 &	2.92$\times$10$^9$ &	1.060 &		49.4\\
UGC 6506 &	15.52 &	2.86$\times$10$^8$ &	0.011 &		11.3\\
NGC 4303 &	12.52 &	9.91$\times$10$^8$ &    (0.180)&		(24.3)\\
\hline
\end{tabular}
\label{satssfr}
\end{table}

Observing time and weather limitations restricted us to spectroscopic observations of only a small subset of our potential satellite galaxies. It is important to establish how representative a subset this is, and to quantify the range of satellite luminosities and star-formation rates to which the validation provided by the present paper applies.  In terms of recession velocity, the 11 host galaxies sample almost the full range of our main satellite search, from 1500 - 6000~km~s$^{-1}$. The main parameters of the 12 satellite galaxies are shown in Table~\ref{satssfr} (the satellite of NGC 4303 was only partially on the CCD field for the H$\alpha$ imaging, so the tabulated quantities are extrapolated and thus less secure than the rest, as indicated by the brackets).  Overall these galaxies appear to resemble typical star-forming field irregular galaxies, or indeed the Magellanic Clouds, in terms of both their $R$-band and their H$\alpha$ luminosities.  The $R$-band and H$\alpha$ luminosity ranges extend from values consistent with small spiral galaxies down to significantly fainter than the SMC. The range of SF rates of our satellites is similar to that of the sample studied by \citet{guti06} (0.01 - 3.7~M$_{\odot}$~yr$^{-1}$).  The most important parameter regarding the reliable detectability of the satellites is their H$\alpha$ equivalent width (EW), since this dictates the strength of the remaining source in the continuum-subtracted image.  The EW values given in Table~\ref{satssfr} are global values, obtained from total $R$-band magnitudes and line fluxes for each satellite; local values on H$\alpha$ peaks within the galaxies will be higher than this. The median EW value is about 16\AA; this is entirely typical of late-type star-forming galaxies, and indeed is somewhat lower than the median values for late-type field galaxies (30 - 38~\AA) found by James et al. (2004).  \citet{guti06} detected significant H$\alpha$ emission from 29 satellite galaxies from the catalogue of \cite{zar97}, with H$\alpha$ EW values ranging from 4 to 441~\AA, and a median value of 23~\AA, which again is slightly larger than that found for our sample.  Thus the imaging method clearly has the sensitivity to detect typical star-forming galaxies over the distance range of interest, and there is no evidence of any bias to, for example, starbursting galaxies with anomalously strong H$\alpha$ emission.

\section{Discussion \& Conclusions}

One obvious limitation of the method is that, while it finds genuine companions of the host galaxies, it only detects strongly star-forming satellites with detectable H$\alpha$ emission. While this may restrict the usefulness of the approach for some applications, it can be argued that, in searching for satellites that may be a source of gas replenishment of the host, the selection of the gas-rich companions is not a disadvantage. The aim of this project using the H$\alpha$ imaging data is to quantify the gas reservoir in satellites available for the replenishment of host galaxy gas content. Gas-rich satellites are most likely to contribute to this reservoir and these are very likely to be star-forming due to tidal interactions with the host or the extended host galaxy H\,{\sc i} halo. The results of this project will be presented in a subsequent paper.

We have obtained optical spectroscopy of 12 faint, apparent emission-line sources, which lie adjacent to nearby spiral galaxies.  These 12 possible companions were initially discovered by imaging through narrow-band H$\alpha$ filters.  The spectroscopy reveals that in every case, the sources detected have genuine line emission, that the line emission detected was H$\alpha$ plus, in most cases, the nearby [N\,{\sc ii}] and/or S\,{\sc ii} lines, and that the velocity differences indicate genuine association between the faint sources and their adjacent spiral galaxies.  Indeed, most and quite possibly all are likely to be bound companions.  Narrow-band imaging thus appears to be an effective method of getting virtually complete samples of at least the star-forming satellite populations of large numbers of galaxies, extending substantially fainter than current surveys based on targeted spectroscopy.  Subsequent papers will analyse the luminosities, radial distributions and star-formation properties of a larger sample of satellites selected using this narrow-band imaging technique.

\section*{Acknowledgments}
This research has made use of the NASA/IPAC Extragalactic Database (NED) which is operated by the Jet Propulsion Laboratory, California Institute of Technology, under contract with the National Aeronautics and Space Administration.
The Isaac Newton Telescope (INT) is operated on the island of La Palma by the Isaac Newton Group in the Spanish Observatorio del Roque de los Muchachos of the Instituto de Astrofisica de Canarias. CFI and PAJ are happy to acknowledge the UK Science and Technology Facilities Council for research studentship and research grant support, respectively.

%\bibliographystyle{mn2e}
%\bibliography{Reference}
\end{document}